\documentstyle[12pt,aasms4]{article}

\received{}
\accepted{}
\lefthead{Pointecouteau et al.}
\righthead{A Sunyaev-Zel'dovich map of the massive core in the luminous X-ray
  cluster RXJ1347-1145}

\begin{document}

%Personal Macros
\def\etal{et al. }
\def\araa{{\it Ann.\ Rev.\ Astron.\ Ap.}}
\def\aplet{{\it Ap.\ Letters}}
\def\aj{{\it Astron.\ J.}}
\def\apj{ApJ}
\def\apjl{{\it ApJ\ (Lett.)}}
\def\apjs{{\it ApJ\ Suppl.}}
\def\aas{{\it Astron.\ Astrophys.\ Suppl.}}
\def\aa{{\it A\&A}}
\def\mnras{{\it MNRAS}}
\def\nature{{\it Nature}}
\def\pasa{{\it Proc.\ Astr.\ Soc.\ Aust.}}
\def\pasp{{\it P.\ A.\ S.\ P.}}
\def\pasj{{\it PASJ}}
\def\pre{{\it Preprint}}
\def\aph{{\it Astro-ph}}
\def\sovlet{{\it Sov. Astron. Lett.}}
\def\adspr{{\it Adv. Space. Res.}}
\def\expas{{\it Experimental Astron.}}
\def\ssr{{\it Space Sci. Rev.}}
\def\inpress{in press.}
\def\souspresse{sous presse.}
\def\inprep{in preparation.}
\def\enprep{en pr\'eparation.}
\def\submit{submitted.}
\def\soumis{soumis.}

\def\ergs{ergs s$^{-1}$}
\def\ergscm{ergs s$^{-1}$ cm$^{-2}$}
\def\pho{ photons cm$^{-2}$ s$^{-1}$ }
\def\phokev{ photons cm$^{-2}$ s$^{-1}$ keV$^{-1}$}
\def\ap{$\approx$ }
\def\ep{$\rm e^\pm$ }
\def\mjyb{mJy/beam }

\def\vp{{ v_{p}}}
\def\te{{T_{e}}}
\def\rc{{r_{c}}}
\def\yc{{y_{c}}}
\def\neu{{n_{e}}}

\def\hzero{{H_{0}}}
\def\qzero{{q_{0}}}
\def\tcmb{{T_{cmb}}}
\def\cd{{C_{dust}}}
\def\td{{T_{dust}}}
\def\nd{{n_{dust}}}
\def\inu{{I_{\nu}}}
\def\fnu{{F_{\nu}}}
\def\bnu{{B_{\nu}}}
\def\mecdeux{{m_{e}c^{2}}}
\def\msol{{M$_{\odot}$}}

\title{A Sunyaev-Zel'dovich map of the massive core in the luminous X-ray
  cluster RXJ1347-1145} 

\author{E. Pointecouteau \altaffilmark{1}, M. Giard \\
Centre d'Etude Spatiale des Rayonnements \\ 
9 av du colonel Roche, BP4346, F-31028 Toulouse cedex 4, France }

\author{A. Benoit \\
Centre de Recherche des Tr\`es Basses Temp\'eratures \\
25 avenue des Martyrs, BP 166 , F-38042, Grenoble Cedex 9, France }

\author{F.X. D\'esert \\
Laboratoire d'Astrophysique de l'Observatoire de Grenoble \\
414 rue de la piscine, F-38041 Grenoble Cedex 9, France }

\author{N. Aghanim, N. Coron, J.M. Lamarre \\
Institut d'Astrophysique Spatiale \\ 
B\^at 121, Universit\'e Paris-Sud, F-91405 Orsay cedex, France}

\author{J. Delabrouille \\
Coll\`ege de France \\
11 place Marcelin Berthelot, F-75231 Paris Cedex 5, France}

\altaffiltext{1}{pointeco@cesr.fr}

\begin{abstract}

We have mapped the Sunyaev-Zel'dovich decrement (hereafter SZ) in the direction
of the  most luminous X-ray cluster known to date, RXJ1347-1145, at
$z=0.451$. This has been achieved with an angular resolution of about 23'' 
using the Diabolo photometer running on the IRAM 30 meter radio telescope. 
We present here a map of the cluster central region at 2.1mm. 
The Comptonization parameter towards the cluster center, 
$\yc=(12.7^{+2.9}_{-3.1})\times 10^{-4}$, corresponds to
the deepest SZ decrement ever observed. Using the gas density distribution 
derived from X-ray data, this measurement implies a gas temperature 
$\te=16.2 \pm 3.8$~keV. The resulting total mass of the cluster is, under 
hydrostatic equilibrium,  $M(r<1\rm{Mpc})=(1.0 \pm 0.3) 
\times 10^{15}$~M$_\odot$ for a corresponding gas fraction 
$f_{gas}(r<1\rm{Mpc})=(19.5 \pm 5.8)\%$.

\end{abstract}


\section{Introduction\label{intro}}

The hot intergalactic gas ($10^{6}$-$10^{8}$K) is, 
with the galaxies themselves and the gravitational effects on background 
objects, one of the tools used
to derive mass distributions within clusters of galaxies. 
It can be detected at X-ray wavelengths via its
bremsstrahlung emission. From submillimeter to 
centimeter wavelengths, the cosmic microwave background (CMB) blackbody
spectrum is distorted in the direction of the cluster 
by the so-called Sunyaev-Zel'dovich
effect (SZ, \cite{sunyaev72}). This characteristic distortion is due
to the inverse Compton scattering of the CMB photons by the intracluster
electrons (see \cite{birkinshaw98} for a detailed review on the SZ effect).

In this letter, we report the SZ measurement of the X-ray cluster RXJ1347-1145
with the ground based Diabolo millimeter instrument. This cluster has been
observed with the ROSAT-PSPC and HRI instruments by Schindler \etal
(\cite{schindler95}, \cite{schindler97}). At a redshift of $z=0.451$, it
appears as the most luminous X-ray cluster ($L_{Bol}=21\times 10^{45}$\ergs)
and so far one of the most massive ($M_{tot}^{Xray}(r<1Mpc)=5.8\times 10^{14}$
\msol). It is also a relatively hot and very dense cluster (temperature: 
$\te=9.3 \pm 1$ keV, central density: $n_{0}=0.094 \pm 0.004$~cm$^{-3}$). 
Optical studies of the
gravitational lensing effects toward RXJ1347-1145 have also been performed
by \cite{fischer97} and \cite{sahu98}. The results have pointed out a
discrepancy between the total mass obtained from the optical and the X-ray
data, with a surface lensing mass toward the core ($r<240$kpc) being one to 
three times higher than the X-ray mass estimates. Because the SZ effect also 
directly probes the projected gas mass, which is not the case for X-ray 
masses, the comparison with SZ measurements might help to discriminate.

In section~\ref{obs}, we describe the Diabolo instrument and our observations
of RXJ1347-1145. The data reduction is explained in section~\ref{red}. In a
fourth part we present the values of the physical parameters we derived from
the data fit.

\section{Observations\label{obs}}

Diabolo is a millimeter photometer which provides an angular resolution of
about 23'' when installed at the focus of the IRAM 30 meter radio
telescope at Pico Veleta (Spain). It uses two wavelengths channels centered 
at about 1.2 and 2.1~mm.
The detectors are bolometers cooled at 0.1 K with an open cycle
$^{4}$He-$^{3}$He dilution refrigerator (Benoit \etal
\cite{benoit98}). Two thermometers 
associated to a heater and a PID digital control system are
used to regulate the temperature of the 0.1 K plate.
There are three adjacent bolometers per channel, arranged in an
equilateral triangle at the focus of the telescope. For a given channel,
each bolometer is coaligned with one bolometer of the second channel, both
looking toward the same sky direction.  
Detections of the SZ effect have already been achieved with Diabolo on 
nearby clusters (A2163, 0016+16, A665) with a single large throughput 
bolometer per channel at 30'' resolution. The experimental setup
is described in D\'esert \etal (\cite{desert98}). The only difference
of the present configuration with the one described in the paper
is the increase of the number of bolometers per wavelength channel
and a slight decrease of the beam FWHM from 30'' to 23''.
With three bolometers at the focus of the telescope, there is no more one
detector on the central optical axis. The 30 meter telescope focus
being of Nasmyth type, the rotation of the field has to be taken into account
in the sky maps reconstruction.

RXJ1347-1145 has been observed in December 1997. 
The center of our observations is the ROSAT-HRI X-ray emission 
center reported by Schindler \etal (\cite{schindler97}):
$\alpha_{2000}=13^{\rm{h}}47^{\rm{m}}31^{\rm{s}}$,
$\delta_{2000}=-11^{\circ}45'11"$.  
The observations have been performed using the wobbling secondary mirror of
the IRAM telescope at a frequency of 1~Hz and with a modulation amplitude of
150''. An elementary observation sequence is a $120''\times 55''$ map in 
right ascension-declination coordinates for a duration of 277 seconds each.
This is obtained using the right ascension
drift provided by the Earth rotation so that the telescope could be 
kept fixed during the measurement. 
This was done to minimize microphonic noises and
electromagnetic influences from the motors driving the IRAM
30m antenna. The map of the cluster is obtained by stepping
in declination between two consecutive lines. The line length is 120''
with a step of 5''. The wobbling is horizontal, thus not aligned with the scan
direction. However, the wobbling amplitude is large enough for the reference
field to be always far out of the cluster. In order to remove systematic
signal drifts that are produced by the antenna environment, we used 
alternatively the positive and negative beam to map the cluster. 
We performed 208 such individual maps on the cluster, for a total duration 
of 16 hours.

Another target of the Diabolo's 1997 run was the direction of
the decrement detected at 8.44~GHz by
Richards \etal (\cite{richards97}). We refer to this source as VLA1312+4237
in the following. Richards \etal 
(\cite{richards97}) measured a flux decrement of $-13.9\pm 3.3\, \mu$Jy in
a 30'' beam. The presence of two quasars in this direction led them claim to
the possible existence of a cluster at a redshift of $z=2.56$. Campos \etal
(\cite{campos98}) have reported the detection of a concentration of
Ly$\alpha$-emitting candidates around the quasars. They argued that the
probability for such a clustering to be random is $5\times10^{-5}$. Our
pointing direction was 
$\alpha=13^{\rm{h}}12^{\rm{m}}17^{\rm{s}}$, $\delta=42^{\circ}37'30''$.
We performed 287 individual maps on this target for a total time
of about 20 hours.

\section{Data reduction and calibration\label{red}}

The reduction procedure includes the following main steps: (i) We remove
cosmic ray impacts. (ii) A synchronous demodulation algorithm is applied taking
into account the wobbling secondary frequency and amplitude. 
(iii) We remove from the
2.1mm bolometer time line the signal which is  correlated with the 1.2mm
bolometer looking at the same sky pixel. This correlated signal is mainly due
to the atmospheric emission, which spectral color is very different from the SZ
effect. (iv) Correction for opacity is done from the bolometer total power
measurements and its calibration by skydips.
(v) To eliminate the low frequency detector noises, a baseline is subtracted 
to each line of the map. The baseline is a 1 degree polynomial. 
It is fitted to 60\% of the data points: 30\% at each
end of the line. (vi) Each map is then resampled on a regular
right-ascension/declination grid, taking into account the field
rotation in the Nasmyth focal plane.
(vii) An average map is computed for each bolometer. Since the weather
conditions were not permanently ideal, the noise quality of the individual maps
is not homogeneous, particularly at 1.2 mm. We thus exclude from 
the average the maps which rms pixel to pixel fluctuation is larger than 1.5 
times the median rms value of all the individual maps. (viii) A single map 
is then produced for each channel (1.2mm and 2.1mm) by coaddition of the 
three bolometer average maps. 

During the run, pointing verifications and mapping of reference sources have
been performed. We have used the planet Mars as a calibration target. 
The apparent angular diameter of Mars was 5'', so that we can consider 
it as a point source with respect to the Diabolo's beam. 
The accuracy of the absolute calibration obtained is of order of 25\% at 1.2mm
and 15\%  at 2.1mm.
Mars observations are also used for the characterization of the Diabolo's
beams. The measured FWHM are 24'' and 22'' at 1.2mm and
2.1mm respectively. Mars has been observed in an azimuth-elevation mapping
mode with a scanning speed which is slower than the natural drift
speed of the cluster observation mode. This later speed is fast enough
compared to the wobbler period to significantly spread the 
signal in the scanning direction (i.e. right ascension). The resulting
beam FWHM for the cluster mode along this direction is 28''. It has 
been experimentally determined
by observation of a quasar lying at about the same declination
as the cluster.

\section{Results\label{resu}} 

The final map of RXJ1347-1145 at 2.1mm is shown on Fig.~\ref{fig1}. 
The X-ray contours have been overplotted. The average right-ascension profile
at 2.1mm is plotted on Fig.~\ref{fig2}. The profile
obtained for the VLA1312+4237 direction using the same data processing,
has been overplotted.
The map and the profiles have been smoothed with a gaussian
filter of 25'' FWHM to maximize the signal to noise ratio. 
The 2.1 mm RXJ1347-1145 map presents a very strong decrement. For a thermal SZ
effect this corresponds to a Comptonization parameter of the order of 
$10^{-3}$. The decrement that we
measure is not centered on the cluster X-ray maximum. We will show in
the analysis that this effect can be explained by the superposition of the SZ
decrement from the intracluster gas and a positive emission from a known radio
source slightly shifted west off the cluster center.

We have no detection for the direction of VLA 1312+4237. 
Our $3 \sigma$ upper limit is  $y < 1.5\times 10^{-4}$. 
This is actually compatible with the decrement measured by
Richards \etal (\cite{richards97}) which translates to a 
central comptonization parameter of the order of 
$7\times 10^{-5}$ for a thermal SZ effect.
If this decrement is in fact due 
to a kinetic SZ effect then we expect a signal at 2.1mm which is equivalent to
a thermal SZ effect of $\yc=1.4\times10^{-4}$, still within
our $3 \sigma$ limit.

Actually, we have used the  VLA1312+4237 data set to obtain a
reliable assessment of the error bars on RXJ1347-1145. 
The individual maps have been averaged over increasing 
durations to evaluate the effective scatter of the average signal
over independent data sets. The maximum duration that could be
checked with this method is about 5 hours, corresponding
to an average of 64 individual maps.  We have checked that for all 
bolometers the rms
pixel noise scales as the square root of the integration time.
The error bars extrapolated from this analysis to longer integration
times are consistent with the error bars derived from the internal
scatter of the data averaged for RXJ1347-1145. The typical
sensitivity reached in the 2.1 mm channel 
is of the order of 1 mJy in a 25'' beam. 

\section{Analysis and discussion\label{ana}}

In the following, we have used for the intracluster gas density
a spherical $\beta$ model with the parameter values derived
from the X-ray analysis of Schindler \etal (\cite{schindler97}):
core radius $\rc=8.4$arcsec (57kpc), $\beta=0.56$, central density
$n_{0} = 0.094$~cm$^{-3}$, and temperature  $\te=9.3$~keV. 
We choose to cut off this distribution at radial distance of 
$r_{cut} = 15 r_c$. We assume the same cosmological parameters too, 
$\hzero=50$km/s/Mpc and $\Omega_{0}=1$ ($\Lambda=0$). With such a model 
the measured SZ skymap reads:

\begin{equation}
I(\bar{\nu},\overrightarrow{\Omega})=\yc\int{\tau (\nu) \, SZ(\nu,\te)d\nu}
\int{P(\overrightarrow{\Omega})L(\overrightarrow{\Omega}-
\overrightarrow{\Omega'})d\overrightarrow{\Omega'}} 
\label{model}
\end{equation}

where $\yc=\displaystyle\frac{k}{\mecdeux}\textstyle
\sigma_{T}\int{\te\neu(r)dl}$, is the
Comptonization parameter towards the cluster center. 
$\neu(r) = n_0 (1+(r/r_c)^2)^{-3\beta/2}$ being the $\beta$ radial
distribution of the gas density. $\tau(\nu)$ is the 
normalized Diabolo band spectral efficiency (given in Desert \etal 
\cite{desert98}). $SZ(\nu,\te)$ is the spectral density of the thermal SZ 
distortion for a unit comptonization parameter, including the relativistic 
weak dependence on $\te$ (see Pointecouteau, Giard \& Barret 
\cite{pointecouteau98}). In fact, for a 9.3 keV cluster, the use of 
relativistic spectra avoids making errors on the SZ flux estimations of 
45\% and 10\% at 1.2 and 2.1 mm respectively. We did not include any
kinetic SZ contribution which is generally weak (\cite{birkinshaw98}).  
 $k$, $m_{e}$, $c$ and $\sigma_{T}$ are respectively the
Bolztman constant, the electron mass, the speed of light and the Thomson 
cross-section. $P(\overrightarrow{\Omega})$ and $L(\overrightarrow{\Omega})$ 
are the normalized angular distributions of the cluster and 
the experimental beam respectively. $P(\overrightarrow{\Omega})$ has no
analytical expression, it is numerically computed by integration of the 
gas density beta profile on the line of sight.

Two radio sources are known from the NRAO VLA Sky Survey 
in the neighborhood of the cluster (\cite{condon98}). 
One, at $(\alpha,\delta)=(13^{\rm{h}}47^{\rm{m}}30.67^{\rm
{s}},-11^{\circ}45^{'}8.6^{''})$, is very close to the cluster center 
and is likely to correspond to the central Cd galaxy. 
Komatsu \etal (\cite{komatsu98}) have compiled observations of this 
radio source at 1.4~GHz, 28.5~GHz and 105~GHz. They
have derived the following power law for the radiosource spectrum: 
$F_{\nu}(band)=(55.7 \pm 1.0)\displaystyle\left(\frac{\nu}{1{GHz}}\right)
^{-0.47\pm 0.02}\textstyle$~mJy. 
So, the extrapolated millimeter flux should be 
$\fnu(1.2\rm{mm})=3.7\pm 0.4$\mjyb and $\fnu(2.1\rm{mm})=4.9\pm 0.5$\mjyb. 

To properly analyze the data, we have performed a realistic 
simulation of the Diabolo observations on the sky map
of the SZ model (Eq.~\ref{model}). 
The whole set of observed individual maps has
been simulated taking into account the 150'' wobbling amplitude
and the proper sky rotation at the Nasmyth focus. 
The simulated data have been processed through the same
pipe-line  as the observed data set to obtain averaged
model maps.

Finally, using this simulated data set, we have simultaneously 
fitted the SZ decrement amplitude and the point
source flux on the 2.1mm profile with $\yc$ and $\fnu(2.1\rm{mm})$ as free
parameters. The best fit parameters are
$\yc=(12.7^{+2.9}_{-3.1})\times10^{-4}$ and
$\fnu(2.1\rm{mm})=6.1^{+4.3}_{-4.8}$\mjyb with a reduced $\chi^{2}$ of
1.3. Results are given at 68\% confidence level. The absolute calibration
error, 25\% and 15\% at 1.2mm and 2.1mm, is not included. $\fnu(2.1\rm{mm})$
is compatible with the value expected from radio
observations. The best fit is overplotted on
the data (see Fig.~\ref{fig2}). It reproduces the asymmetric profile. This
asymmetry is due to the point 
source contribution which fills part of the SZ decrement. 

In a second time, we have fixed the radio point source flux at the expected
value deduced from Komatsu \etal (\cite{komatsu98}), $\fnu(2.1mm)=4.9$\mjyb, 
and we have fitted with a maximum likelihood method both the central 
Comptonization parameter $\yc$, and the angular core radius $\theta_{c}$. 
We have found $\yc=(13.2^{+0.2}_{-2.6})\times10^{-4}$ and
$\theta_{c}=7.2^{+7.3}_{-7.2}$arcsec with a reduced $\chi^{2}$ of 1.2. The
Comptonisation parameter value is consistent with the previous one. The angular
core radius is consistent with the X-ray value within the 68\% confidence
level.

\subsection{Conclusion\label{discu}}

We confirm through our SZ detection that RXJ1347-1145 is an extremely massive
and hot cluster. We have measured the deepest SZ effect ever observed. 
It corresponds to a very large Comptonization parameter,
$\yc=(12.7^{+2.9}_{-3.1})\times10^{-4}$. This is almost twice the value
expected from the X-ray data, $y_{Xray}=(7.3 \pm 0.7)\times10^{-4}$
if we use the cluster gas parameters derived by Schindler et al. 
(\cite{schindler97}). Although our result points to a mass higher than the
X-ray mass, as is the case for gravitational lenses measurements, 
the uncertainties do not allow to firmly conclude to a discrepancy.
The X-ray flux toward the cluster center is actually dominated by the very
strong cooling flow in the core. The average temperature of the gas
which contributes to the SZ effect is thus likely to be higher than
the temperature derived from the X-ray data, $\te = 9.3 \pm 1$ keV.
The gas temperature needed to produce the thermal SZ effect we have
observed is $\te=16.2 \pm 3.8$~keV, assuming all other parameters
are kept unchanged.
In a re-analysis which takes into account the heterogeneity of the 
cluster, Allen and Fabian (\cite{allen98}) have actually derived for this
cluster a very high gas temperature: $\te = 26.4^{+7.8}_{-12.3}$ keV,
which is indeed consistent with our measurement.
Under the hypothesis of hydrostatic equilibrium, a higher 
gas temperature implies a higher total cluster mass, thus
decreasing the gas fraction if all other cluster parameters
are kept unchanged. For $\te=16.2 \pm 3.8$~keV the total mass of
RXJ1347-1145 within 1~Mpc is considerable, 
$M_{tot}(r<1\rm{Mpc})=(1.0 \pm 0.3) \times 10^{15}$~M$_{\odot}$ 
and the corresponding gas fraction is $f_{gas}(r<1\rm{Mpc})=(19.5 \pm 5.8)\%$.

\acknowledgments
We are very grateful to the IRAM staff at pico Veleta for their help during
observations. We thank Laurent Ravera for very useful comments during the data
analysis phase. Diabolo is supported by the Programme National de
Cosmologie, Institut National pour les Sciences de l'Univers,
Minist\`ere de l'Education National de l'Enseignement sup\'erieur et de
la Recherche, CESR, CRTBT, IAS-Orsay and LAOG. We thank the anonymous referee 
for numerous comments and corrections which allowed us to considerably
improve the paper.

\clearpage

\centerline{\bf FIGURE LEGENDS}

\figcaption[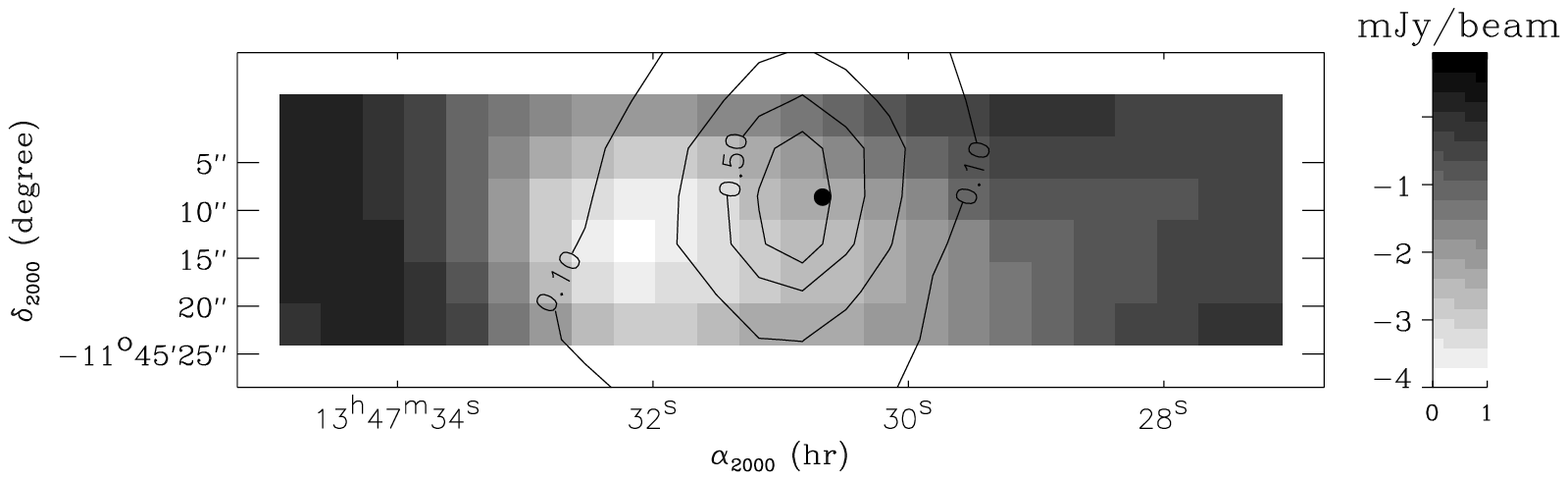]{2.1mm map of the RXJ1347-1145 central region 
  obtained with Diabolo. The map has been smoothed with a 25'' FWHM gaussian 
  filter. The 1$\sigma$ noise is about 1\mjyb. The X-ray contours have been
  overplotted. The big dot indicates the radio source position.
\label{fig1} }

\figcaption[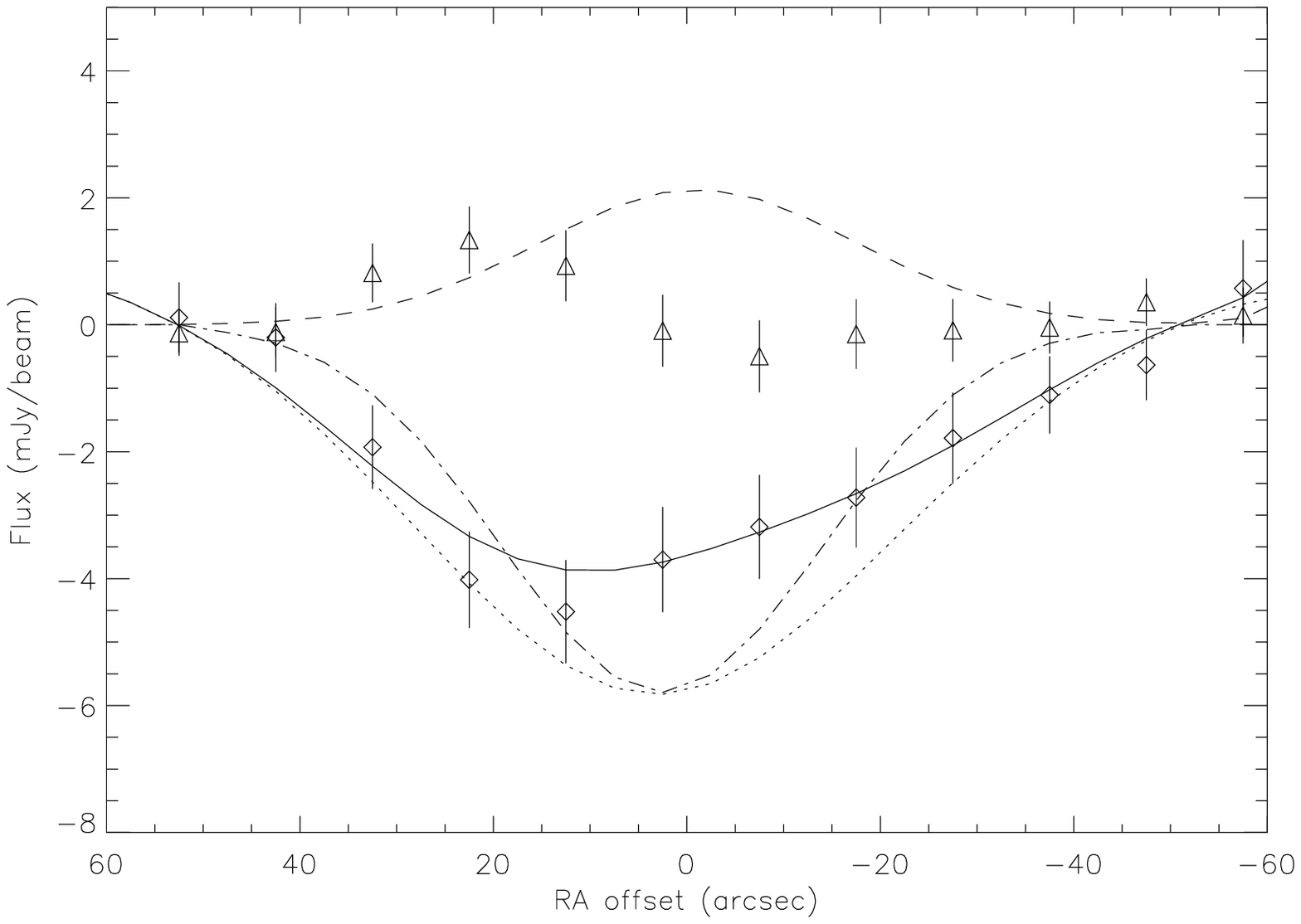]{2.1mm right ascension profile of RXJ1347-1145 
  (data smoothed with a 25'' FWHM gaussian). 
  The data points have been plotted (diamond)  with their
  1$\sigma$ error bars. 
  The best fit model (solid line) combines a SZ component (dotted
  line) plus a point source component (dashed line). 
  The dotted-dashed line draws a point source
  profile with the same amplitude as the SZ effect. It shows that the
  decrement we observed is more extended than a point source. The 
  VLA1312+4237 average profile (no detection, see text) 
  is shown with triangles.  
 \label{fig2}}

\clearpage
\plotone{pointeco_fig1.eps}
{\centerline {Figure \ref{fig1}}}

\clearpage

\plotone{pointeco_fig2.eps}
{\centerline {Figure \ref{fig2}}}

\end{document}